\RequirePackage{fix-cm}
\documentclass[smallextended]{svjour3}       
\smartqed  
\usepackage{graphicx}
\usepackage{todonotes}
\usepackage{booktabs} 
\usepackage{amssymb}
\usepackage{enumitem}
\usepackage{tikz}
\usepackage[T1]{fontenc}
\newcommand{\citet}{\cite}
\usepackage[numbers]{natbib}
\usepackage{hyperref}
\usepackage{fancyvrb}
\definecolor{olivegreen}{HTML}{3f7d31}
\hypersetup{
    colorlinks,
    linkcolor={blue!50!black},
    citecolor={blue!50!black},
    urlcolor={blue!50!black}
}
\newcommand{\up}{$\blacktriangle$~}
\newcommand{\dn}{$\blacktriangledown$~}
\begin{document}

\title{Overcoming Low-Utility Facets for \\ Complex Answer Retrieval
\thanks{This is an extended version of ``\textit{Characterizing Question Facets for Complex Answer Retrieval}. TREC 2018.''~\cite{MacAvaney2018} Extensions include the use of knowledge graphs to inform model, and an extensive analysis.}
}


\author{Sean MacAvaney \and Andrew Yates \and Arman Cohan \and Luca Soldaini \and Kai Hui \and Nazli Goharian \and Ophir Frieder}


\institute{S. MacAvaney \at
              Information Retrieval Lab, Georgetown University \\
              3700 O St NW,
              Washington, DC 20057, United States\\
              Tel.: +1-202-687-5874\\
              Fax: +1-202-687-7822\\
              \email{sean@ir.cs.georgetown.edu}
           \and
           A. Yates \at Max Planck Institute for Informatics
           \and
           A. Cohan \and L. Soldaini \at Information Retrieval Lab, Georgetown University
           \and
           K. Hui \at SAP SE
           \and
           N. Goharian \and O. Frieder \at Information Retrieval Lab, Georgetown University
}

\date{Received: date / Accepted: date}

\maketitle

\begin{abstract}
Many questions cannot be answered simply; their answers must include numerous nuanced details and additional context. Complex Answer Retrieval (CAR) is the retrieval of answers to such questions. In their simplest form, these questions are constructed from a topic entity (e.g., `cheese') and a facet (e.g., `health effects'). While topic matching has been thoroughly explored, we observe that some facets use general language that is unlikely to appear verbatim in answers. We call these low-utility facets. In this work, we present an approach to CAR that identifies and addresses low-utility facets. We propose two estimators of facet utility. These include exploiting the hierarchical structure of CAR queries and using facet frequency information from training data. To improve the retrieval performance on low-utility headings, we also include entity similarity scores using knowledge graph embeddings. We apply our approaches to a leading neural ranking technique, and evaluate using the TREC CAR dataset. We find that our approach perform significantly better than the unmodified neural ranker and other leading CAR techniques. We also provide a detailed analysis of our results, and verify that low-utility facets are indeed more difficult to match, and that our approach improves the performance for these difficult queries.

\end{abstract}

\section{Introduction} 
\label{sec.introduction}

It is common to use search technologies to find answers to questions. While considerable work has been done on questions that have simple factoid answers, less focus has been given to questions with complex answers. Here we define complex answers as answers that include many details or additional context (making complex answers different than factoid answers). Complex Answer Retrieval (CAR) is the process of finding answers to questions that have complex answers~\cite{dietz2017car}. Note that CAR questions are not necessarily complex. A question as simple as \textit{`Is cheese healthy?'} requires a complex answer: a detailed and nuanced description of positive and negative health effects of cheese consumption is required to satisfy the information need. In contrast, a question such as \textit{`How much Mozzarella cheese do I need to eat to satisfy my daily requirement of calcium?'} is a complex question with a simple factoid answer.\footnote{This question is complex because it involves advanced reasoning that goes beyond what is typically captured by a knowledge graph.}

Since complex answers include details and context, they often are formulated as paragraphs of text. Thus, CAR is framed as paragraph retrieval for a question that has a complex answer. We propose two complementary approaches to identify question facts that are difficult to match, and one knowledge graph-based approach to assist in the matching of questions with difficult facets. Due to prior showing that neural models are more effective for CAR than traditional IR techniques~\cite{DBLP:journals/corr/NanniMMD17}, we build our approaches into a leading neural ranker to validate our methods.

\begin{figure*}[!b]
\hrule\vspace{1em}
\begin{Verbatim}[commandchars=\\\{\}]
Question: Is cheese healthy?
Topic: Cheese
Facet: Health effects

\textcolor{olivegreen}{Answer 1: Although nutrient levels vary by type of \underline{cheese}, most \underline{cheese} are a}
\textcolor{olivegreen}{          rich source of calcium, protein, and sodium. Because the primary}
\textcolor{olivegreen}{          ingredient of cheese is milk, most of most of the nutritional...}
\textcolor{olivegreen}{Answer 2: Consumption of foods that are high in saturated fat such as \underline{cheese}}
\textcolor{olivegreen}{          are linked with an increase risk of cardiovascular disease.}
\textcolor{olivegreen}{          This is due to adverse +underline[effects] to blood lipids and circulating...}
\textcolor{olivegreen}{Answer 3: \underline{Cheese} has the potential for promoting the growth of Listeria}
\textcolor{olivegreen}{          bacteria. This can cause serious infection in an infant and pregnant}
\textcolor{olivegreen}{          woman and can be transmitted to her infant in utero or after...}
\textcolor{red}{Answer 4: Wisconsin is known as "America's Dairyland" because it is one of}
\textcolor{red}{          the nation's leading dairy producers, particularly famous for its}
\textcolor{red}{          \underline{cheese}. Manufacturing, \underline{health} care, information technology, and...}
\end{Verbatim}
\hrule
\caption{Hypothetical CAR query with three relevant answers (1-3) and one non-relevant answer (4). Term matches from the topic and facet are underlined.}
\label{fig:ex-query}
\end{figure*}

CAR queries can be broken down into two components: the topic and facet. The topic is the main entity of the question. For the query \textit{`Is cheese healthy?'}, the topic is the entity `Cheese' (see Figure~\ref{fig:ex-query}). All answers to the question must be about this entity, otherwise the answer is not valid. The facet is the particular detail about which the question inquires. In the example, the facet can be described as `Health effects'. For an answer to be considered valid to the question, it must refer to health effects of cheese---information such as nutrition, health risks, etc. These answers can come from multiple sources. For instance, an article about cardiovascular disease may claim that diets containing foods such as cheese that are high in saturated fat increase one's risk of heart disease (Answer 2 in Figure~\ref{fig:ex-query}). Such a paragraph would be valuable to include in a complete answer because it contains contextual information about why cheese consumption can increase one's risk of heart disease. Thus, it makes sense to frame CAR as retrieval of paragraphs of text from authoritative sources given a topic and facet. CAR queries with various facts about a single topic can be combined to produce a detailed article about a topic.

A straightforward approach for CAR is to simply use an existing IR technique as-is, concatenating the topic and facet information to build the query. However, such an approach is limited by several factors specific to CAR. We observe that facets are not necessarily mentioned verbatim in relevant paragraphs. In Figure~\ref{fig:ex-query}, the relevant answers (1-3) never include `health' because the context is clear from the other entities mentioned in the answer. On the other hand, non-relevant answers sometimes do include the facet term (e.g., answer 4). An effective CAR engine needs to account for this. Not all facets exhibit this general behavior. Let's consider another query: \textit{``What is the effect of curdling in cheese production?''} (topic: cheese, facet: curdling). Since `curdling' is not a general-language term, relevant answers probably need to use the term itself; related entity mentions are probably inadequate and would result in confusing text. We refer to this distinction as \textit{facet utility}: high-utility facets use language that is specific to the topic and can be found directly in relevant answers (e.g., `curdling'), whereas low-utility facets use general language and requires additional domain knowledge to identify relevant paragraphs (e.g., `health effects').

Given these observations, we propose a two-pronged approach to CAR. First, we attempt to predict facet utility. For this, we use both structural information about the query itself, and corpus statistics about how frequently facets are used. Then, to better accommodate low-utility facets, we utilize entity mentions in the candidate answer. To this end, we construct a knowledge graph embeddings using a training corpus, and measure distances between the topic entity and the entity mentions. We incorporate both the facet utility estimators and the entity scores into a neural ranking model, and use the model to retrieve complex answers. These approaches yield significant improvements over the unmodified neural ranker, and up to 26\% improvement over the next best approach. We then provide a detailed analysis of our results, which shows that, indeed, low-utility facets are indeed more difficult to match, and that our approach improves these results. As one of the first comprehensive works on CAR, we expect these insights to guide future work in CAR.

\section{Background and Related Work}\label{sec.background}

\subsection{Complex Answer Retrieval}

\begin{figure}
\centering
\small
\scalebox{1.3}{
\begin{tikzpicture}[xscale=1,yscale=1]
\node(cheese){\includegraphics[scale=0.35]{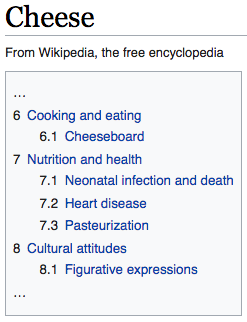}};
\draw [<-, thick] (-0.2,1.77) -- (1.8,1.77) node[fill=white, pos=1, anchor=west]{topic entity (title)};
\draw [<-, thick] (0.35,0.03) -- (1.8,0.03) node[fill=white, pos=1, anchor=west]{facet (heading)};
\end{tikzpicture}
}
\caption{Headings found in the Wikipedia article \textit{Cheese}, including labels for the question ``\textit{Is cheese healthy?}''. Adapted from \url{https://en.wikipedia.org/wiki/Cheese}.}\label{fig.cheese}
\end{figure}

Complex answer retrieval is a new area of research in IR. In 2017, the TREC conference ran a new shared task focused on CAR~\cite{dietz2017car}. The goal of this task is to rank answer paragraphs corresponding to a complex question. The shared task frames CAR in terms of Wikipedia content generation. This is appropriate because the editorial guidelines and strong role of moderators makes Wikipedia an authoritative source of information~\cite{heilman2015wikipedia}. Paragraphs from articles meeting topic criteria\footnote{E.g., templates, talk pages, portals, lists, references, and pages representing people, organizations, music, books, and others are discarded~\cite{dietz2017car}.} are selected as a source of answers for retrieval. The task goes one step farther by asserting that Wikipedia is also a good source of CAR queries and relevance judgments. CAR queries are formed from article titles (the query topic, an entity), and headings (query facets of that particular topic). Furthermore, paragraphs found under each heading are treated as ``automatically'' relevant to that particular topic, yielding a large amount of training data~\cite{dietz2017cardata}. This makes it practical to train data-hungry neural IR models for CAR. Figure~\ref{fig.cheese} shows an article's heading hierarchy, including the question topic and an example facet.

Due to the hierarchal nature of headings in Wikipedia, CAR queries include multiple headings. This is important to provide adequate context for the answers. For instance, in Figure~\ref{fig.cheese}, heading 7.3 refers specially to the pasteurization of cheese as it relates to nutrition and health. For simplicity, we treat the article title as the first element in a list of headings that represents each query. Thus, a CAR query is a sequence of headings. In the remainder of this manuscript, we use `\guillemotright' to separate the heading components of a query. The root heading is the \textit{title} and represents the topic entity of the query. The leaf heading represents the most specific topic; we call this the \textit{main heading}. \textit{Intermediate headings} are any headings between the title and main heading provide context for the main heading. Table~\ref{tab:ex-queries} provides example queries using this terminology. Note that a given query does not necessarily have any intermediate headings, and that a main heading in one query (\textit{Cheese \guillemotright{} Nutrition and health}) can appear as an intermediate heading in another query (\textit{Cheese \guillemotright{} Nutrition and health \guillemotright{} Pasteurization}).

\begin{table*}
\centering
\setlength{\tabcolsep}{0.5em}
\caption{Example CAR queries from Wikipedia by heading position. Some queries have no intermediate headings.}\label{tab:ex-queries}
\begin{tabular}{lllll}
\toprule
Title & & Intermediate Heading(s) & & Main Heading \\
\midrule
Cheese &\guillemotright{}& \textit{(none)} &\guillemotright{}& Nutrition and health\\
Green sea turtle & \guillemotright & Ecology and behavior & \guillemotright & Life cycle \\
History of the United States & \guillemotright & 20th Century & \guillemotright & Imperialism \\
Disturbance (ecology) & \guillemotright & \textit{(none)} & \guillemotright & Cyclic disturbance \\
Medical tourism & \guillemotright & Destinations \guillemotright{} Europe & \guillemotright & Finland \\
\bottomrule
\end{tabular}
\end{table*}

Existing work in CAR is relatively limited. Prior to TREC, \citet{DBLP:journals/corr/NanniMMD17} investigated baseline approaches using the automatic relevance judgments (assuming only paragraphs found under a heading are relevant to that query). Their approaches include BM25, cosine similarity (both with TF-IDF vectors and word embeddings), a baseline learning-to-rank approach, query expansion, and a deep neural model. They found that the deep neural model Duet~\cite{mitra2017learning} outperforms the others.

The 2017 TREC CAR task inspired several new approaches to CAR. In this work, we extend our submission to TREC and some follow-up analysis~\cite{MacAvaney2017,MacAvaney2018}. Another top-performing submission uses a Sequential Dependence Model (SDM~\cite{Metzler2005AMR}) for answer retrieval~\cite{cuis2017}. They modified the SDM for CAR by considering ordered ngrams that occur within an individual heading component, and unordered ngrams to terms in different headings. This results in a bias toward matching partial phrases that appear in individual headings, and reduces processing time for long queries. Another approach uses a Siamese attention network~\cite{utd2017}. Features incorporated into this network include abbreviated entity names and lead paragraph entity mentions from DBPedia~\cite{Auer2007DBpediaAN}. Yet another approach uses reinforcement learning query reformulation for CAR~\cite{nyudl2017,Nogueira2017TaskOrientedQR}. The query reformulation step helps the model add terms that make up for low-utility facets. Our approach differs from these by (1) explicitly modeling facet utility to predict which headings are unlikely to appear in relevant documents, and (2) incorporating contextualized entity relatedness measures to improve performance on low-utility facets.

\subsection{Neural IR Models}

Recent advances in neural IR have produced promising results in the ad-hoc domain, making them a potentially useful tool for CAR ranking. Early work on neural IR models focused on building dense representations of texts that are capable of being compared to measure relevance (e.g., Deep Structured Semantic Model (DSSM)~\cite{huang2013learning}). These \textit{representation-focused} methods are akin to traditional vector space model approaches. While scalable, representation-focused methods suffer from loss of term locality information, making certain types of query term matching difficult.

\begin{figure*}
\centering
\includegraphics[scale=1.1]{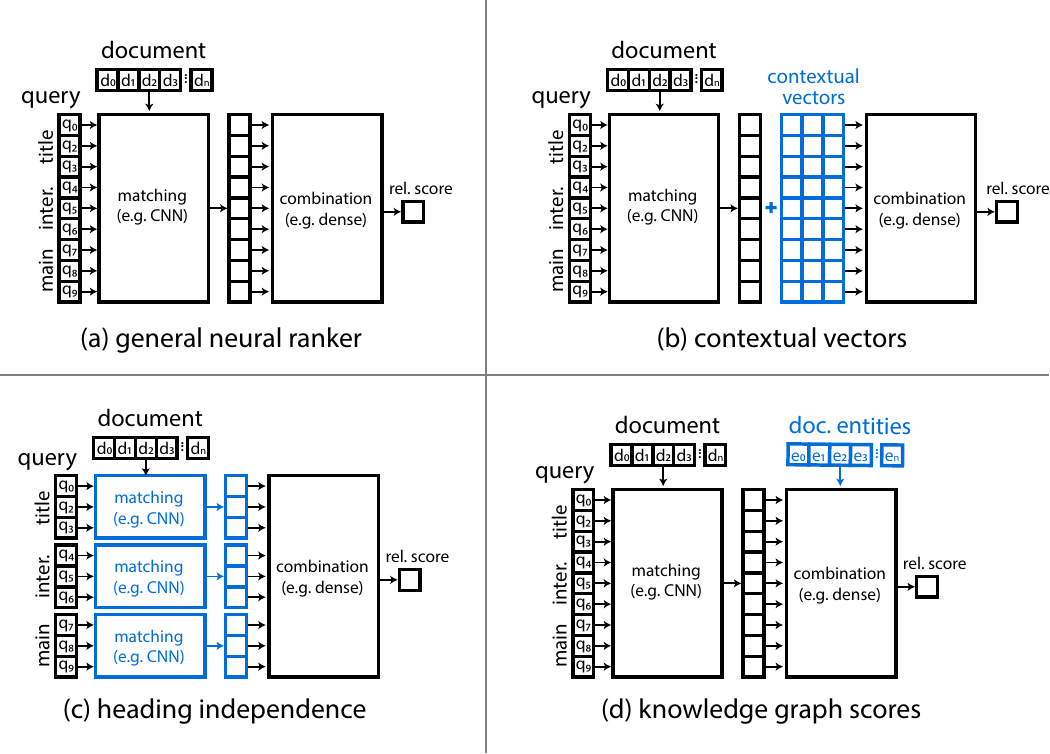}
\caption{(a) General ranking architecture, with matching and combination phases (unmodified). (b) Modified architecture, including contextual vectors for combination. (c) Modified architecture, splitting for heading independence. (d) Modified architecture, including knowledge graph scores in the combination layer.}\label{fig.model}
\end{figure*}

More recently, work has shifted to \textit{interaction-focused} methods that model how query terms appear in relevant documents directly, i.e. what patterns of query terms are typical in relevant documents. These approaches typically use a two-phase approach (see Figure~\ref{fig.model}a). The first phase finds term matches for a query-document pair, often using a convolutional neural network (CNN) and result pooling. The CNN often iterates over a query-document similarity matrix of pretrained word embedding cosine similarities. The second phase combines the results for each query term using, for instance, a dense feedforward network. The combination results in a relevance score for the query-document pair, which is then used for ranking. For instance, the prominent PACRR model~\cite{Hui2017PACRRAP} implements the matching layer with a CNN that is max pooled over filters, then $k$-max pooled over query terms to pick out the most salient signals. Then, the results are combined with a dense combination phase. Generally, these neural models have been shown to be competitive with conventional ranking techniques due to their ability to learn matching patterns (e.g., bigrams, trigrams, skipgrams) from the training data, and combine the results in a meaningful way. Other prominent interaction-focused architectures include DRMM~\cite{guo2016deep}, MatchPyramid~\cite{Pang2016ASO}, K-NRM~\cite{xiong2017neural}, Duet~\cite{mitra2017learning}, DeepRank~\cite{Pang2017DeepRankAN}, Conv-KNRM~\cite{Dai2018ConvolutionalNN}, and Co-PACRR~\cite{Hui2018CoPACRRAC}. In this work, we choose to modify a neural ranking architecture because preliminary work has shown them to be more effective than conventional approaches out-of-the-box~\cite{DBLP:journals/corr/NanniMMD17} and they can easily incorporate new techniques and information specific to this task. Specifically, by default no rankers consider facet utility, and in this work we show how this can be incorporated into a neural ranking model.


\subsection{Wikipedia and Knowledge Graphs}

Since knowledge graphs encode which entities are related to one another, they may be valuable when identifying paragraphs with low-utility facets. There have been many efforts to organize Wikipedia information into a knowledge graph. Prominent efforts have been the DBPedia~\cite{Auer2007DBpediaAN} and Freebase~\cite{Bollacker2008FreebaseAC}. Researchers have found that query expansion using knowledge bases can improve ad-hoc retrieval~\cite{dalton2014entity,xiong2015fbexpansion}. Others have investigated how to include knowledge graph features directly in learning-to-rank approaches~\cite{Schuhmacher2015RankingEF}.
Other efforts have been in how to train embeddings for entities and relations in knowledge graphs. Prominent approaches include translational embeddings (TransE)~\cite{Bordes2013TranslatingEF}, translational hyperplane embeddings (TransH)~\cite{Wang2014KnowledgeGE}, and holographic embeddings (HolE)~\cite{Nickel2016HolographicEO}. Knowledge graphs have also been employed extensively for general question answering tasks~\cite{yih2015semantic,Singh2012EntityBQ}. In this work, we observe that entity similarity act as a signal for answers that are otherwise difficult to match (i.e., when facets are low-utility). We build a knowledge graph from Wikipedia articles using entity mentions and heading labels. We then train embeddings, and use embedding similarities when ranking answers.

\section{Method}\label{sec.method}

As mentioned in Section~\ref{sec.background}, complex answer retrieval (CAR) is a new IR task focused on retrieving complex answers to questions that include a topic and facet. Paragraphs from authoritative resources (e.g., Wikipedia) are considered partial answers for these questions. Identifying and overcoming low-utility facets is a central challenge to CAR~\cite{MacAvaney2017}. Here, we propose approaches based on the Wikipedia-focused CAR problem:

\begin{itemize}
\item \textit{High-utility facets.} We find that high-utility facets correspond to headings that relate to specific details of an article's topic: \textit{topical headings}. Thus, a given topical heading is unlikely to appear in most articles. However, we predict that terms found in topical headings are \textit{more} likely to appear in relevant paragraphs because people are less likely to have the existing knowledge to determine meaning from the mentioned entities. Since the title of an article is the article's topic, it is necessarily topical.

\item \textit{Low-utility facets.} We identify that low-utility facets often correspond to structural headings in Wikipedia. These headings provide coherence across articles by enforcing a standard document structure. As a result, these headings occur frequently. Furthermore, they often appear as intermediate headings by organizing topical headings below them. Since the terminology is necessarily more general, we predict that terms in structural headings are less likely to appear in relevant paragraphs because readers can figure out the context by related entities. Thus, we propose including additional entity similarity information to accommodate these cases.
\end{itemize}

Recall that an interaction-focused neural ranker (Figure~\ref{fig.model}a) involves two phases: a matching phase that identifies places in the document in which query terms are used (e.g., convolution), and a combination phase in which the scores for each query term are combined to produce a final relevance score (e.g., dense). In the remainder of this section, we present two approaches to inform an arbitrary interaction-focused neural ranker of facet utility (Section~\ref{sec:meth:context}-\ref{sec:meth:indep}, Figure~\ref{fig.model}b-c), and one approach to include entity similarity signals to inform the model of relevance for low-utility facets using knowledge base embedding similarity scores (Section~\ref{sec:meth-kg}, Figure~\ref{fig.model}d). Finally, we describe our implementation strategy for a concrete neural ranker (PACRR, Section~\ref{sec:meth-pacrr}).

\subsection{Contextual vectors}\label{sec:meth:context}

Many information retrieval approaches use the simple (yet powerful) term IDF value as a signal for query term importance. Neural models incorporate this signal, too. For instance, DRMM~\cite{guo2016deep} and PACRR~\cite{Hui2017PACRRAP} use an IDF vector in the combination phase. We generalize this approach by allowing an arbitrary number of contextual vectors to be included in the model alongside term interaction signals (Figure~\ref{fig.model}b). Here, we use contextual vectors to provide an estimation of facet utility, understanding that the models can learn how to use these values when making relevance judgments. We propose two such estimators: \textit{heading position}~(HP) and \textit{heading frequency}~(HF).

\begin{table*}
\setlength\tabcolsep{0.5em}
\caption{Sample contextual vectors for ``\textit{green sea turtle \guillemotright{} ecology and behavior \guillemotright{} life cycle}''.}\label{tab:vecs}
\centering
\begin{tabular}{l|ccc|ccc|cc}
\toprule
 & green & sea & turtle & ecology & and & behavior & life & cycle \\
\midrule
position\_title    & 1 & 1 & 1 & 0 & 0 & 0 & 0 & 0 \\
position\_inter    & 0 & 0 & 0 & 1 & 1 & 1 & 0 & 0 \\
position\_main     & 0 & 0 & 0 & 0 & 0 & 0 & 1 & 1  \\
\midrule
heading\_frequency & 0 & 0 & 0 & 3 & 3 & 3 & 3 & 3 \\
\bottomrule
\end{tabular}
\end{table*}

\paragraph{Heading position} Recall that CAR queries consist of multiple components: the topic and facet. When framed as Wikipedia content generation, CAR queries are broken down further into a list of sub-queries, where the first represents the topic, followed by intermediate headings and the main heading. When estimating heading utility, the position of the heading in the list intuitively provides a signal of heading utility. For instance, the title is necessarily topical; it is the topic. Furthermore, any intermediate headings are likely structural because they provide categorizations for the main heading. Finally, the main heading may be either topical or structural; its position inherently tells the model nothing about whether or not to expect the term to appear. Thus, we encode three vectors to encode query term position information: one that indicates if the term exists in the title, and intermediate heading, or the main heading. An example of these vectors is given in Table~\ref{tab:vecs}.

The benefits of including heading position information can extend beyond simply distinguishing whether a term is likely in a topical or structural heading. For instance, the topic of a question may be abbreviated in relevant paragraphs to avoid excessive repetition. By including heading position information in this way, the model can distinguish when certain matching patterns are important to capture.

\paragraph{Heading frequency} Another possible estimator of heading utility is the frequency that a heading appears in a sufficiently representative dataset. For instance, because structural headings such as \textit{``Nutrition and health''} are general and can accommodate a variety of topics, you would expect to find the heading in many food-related articles. Indeed, the heading appears in Wikipedia articles such as \textit{Cheese}, \textit{Beef}, \textit{Raisin}, and others. On the other hand, one would expect topical headings to use less general language, and thus be less likely to appear in other articles. For instance the heading \textit{`After the Acts of Union of 1707'} only appears in the article \textit{`United Kingdom'}.

To represent this value, we use the document frequency of each heading: the \textit{heading frequency}. Thus, frequent headings (e.g., `History') have a large heading frequency value, and infrequent headings (e.g., `After the Acts of Union of 1707') have a low value. For heading matching, we require a complete, case-insensitive match of the text. Thus, `Health effects' and `Health Effects' are considered the same heading (capitalization), but not `Health effects' and `Health' (substring) or `Health affects' (typographical error). To group similarly-frequent headings together, we stratify the heading frequency value when used as a contextual vector in a range of $[0,3]$. In pilot studies, we found effective breakpoints to be the 60th percentile (approximately the cutoff for headings that only appear a few times, such as \textit{Red Hot Chili Peppers}), 90th percentile (approximately the cutoff of moderately frequent headings, such as \textit{Finland}), and 99th percentile (approximately the cutoff of frequent headings, such as \textit{Family and personal life}). An example of this vector is given in Table~\ref{tab:vecs}.

When the model encounters the heading frequency value, we expect it to learn to value high-frequency headings less than low-frequency headings (i.e., assign a negative weight). This results in similar behavior to IDF, but with the important distinction that it operates over entire headings, rather than terms. For instance, most terms in \textit{`After the Acts of Union of 1707'} have a low IDF (i.e., they appear frequently), but as a whole heading, it appears infrequently and is likely important to match.

\paragraph{Combining contextual vectors}
The two contextual vectors capture different notions of heading utility. Heading position contextual vectors are able to discriminate otherwise identical headings based on where they appear in the query. This can be valuable in some situations: for instance, `Imperialism' and `Finland' in Table~\ref{tab:ex-queries} appear as main headings, but they could also appear as the topic of other queries. Heading frequency contextual vectors treat headings with the same text identically, regardless of their position in the query. However, their power comes from informing the model about the nature of the specific heading, and can help identify the utility of main headings (which are otherwise ambiguous), or headings that do not match the typical characteristics in other heading positions. By including both vectors, the model should be able to learn a better sense of heading utility by combining the two notions.

\subsection{Heading independence}\label{sec:meth:indep}

Although contextual vectors can help inform a trained ranker about which headings are structural and topical, they cannot directly affect which types of interactions are identified and how these interactions are scored because this occurs earlier in the pipeline of the neural model considered (matching phase). We hypothesize that heading utility can actually affect which signals are important. For instance, a low-utility structural heading might be more tolerant to weak term similarities (i.e., `colonial' or `ancient' might be acceptable matches for `History'), or a high-utility topical heading might have stricter requirements for maintaining the order of terms. Thus, we propose a general approach to modify a neural ranking architecture to adjust matching based on heading utility: \textit{heading independence}.

Heading independence involves splitting the matching phase of a neural ranker into multiple segments, each responsible for processing a segment of the query. Here, we split the query by heading position: title, intermediate, and main headings (see Figure~\ref{fig.model}c). The processing of each heading position is independent of the others, so a different set of parameters can be learned for each component. The results of the matching layers are then concatenated and sent to the combination layer of the unmodified model (e.g., a dense layer) for the calculation of the final relevance score.

\begin{figure*}[b!]
\hrule\vspace{1em}
\begin{verbatim}
Q1: Cheese (H1) > Nutrition (H3) and (H3) health (H3)
Q2: Medical (H1) tourism (H1) > Destinations (H2) > Europe (H2) > Finland (H3)

             Unaligned                       Aligned
    --------------------------     --------------------------
Q1: H1 H3 H3 H3 __ __ __ __ __     H1 __ __ __ __ __ H3 H3 H3
Q2: H1 H1 H2 H2 H3 __ __ __ __     H1 H1 __ H2 H2 __ H3 __ __
\end{verbatim}
\hrule
\caption{Example term alignment benefits of using heading independence with two sample queries. H1: query title. H2: intermediate heading. H3: main heading. \_\_: padding.}
\label{fig:ex-alignment}
\end{figure*}

This approach makes intuitive sense because one would expect the title to interact differently in the document than a structural or topical heading. For instance, abbreviated versions of the topic are often used to improve readability. Furthermore, this approach enforces alignment of headings during combination (see Figure~\ref{fig:ex-alignment}). When all query terms are simply concatenated, the alignment of each query position will change among queries due to differences in query length. Thus, the model will need to learn and accommodate multiple tendencies in a wider variety of positions of the query, rather in just certain areas.

\subsection{Knowledge graph}\label{sec:meth-kg}

Although interaction-focused neural rankers often employ word embedding similarities, word embeddings do not always capture similarities within contexts. This results in individual term being close to many topically-similar terms~\cite{levy2014dependency}. Within the context of CAR, we are only concerned with topics that are similar given a particular context (i.e., the question facet). Furthermore, we know that the most salient contextual information comes from entity mentions. Since article titles also correspond to entities, and headings can be considered the context, we use the following approach employing knowledge graph embeddings.

\begin{figure}
\centering
\includegraphics[scale=0.4]{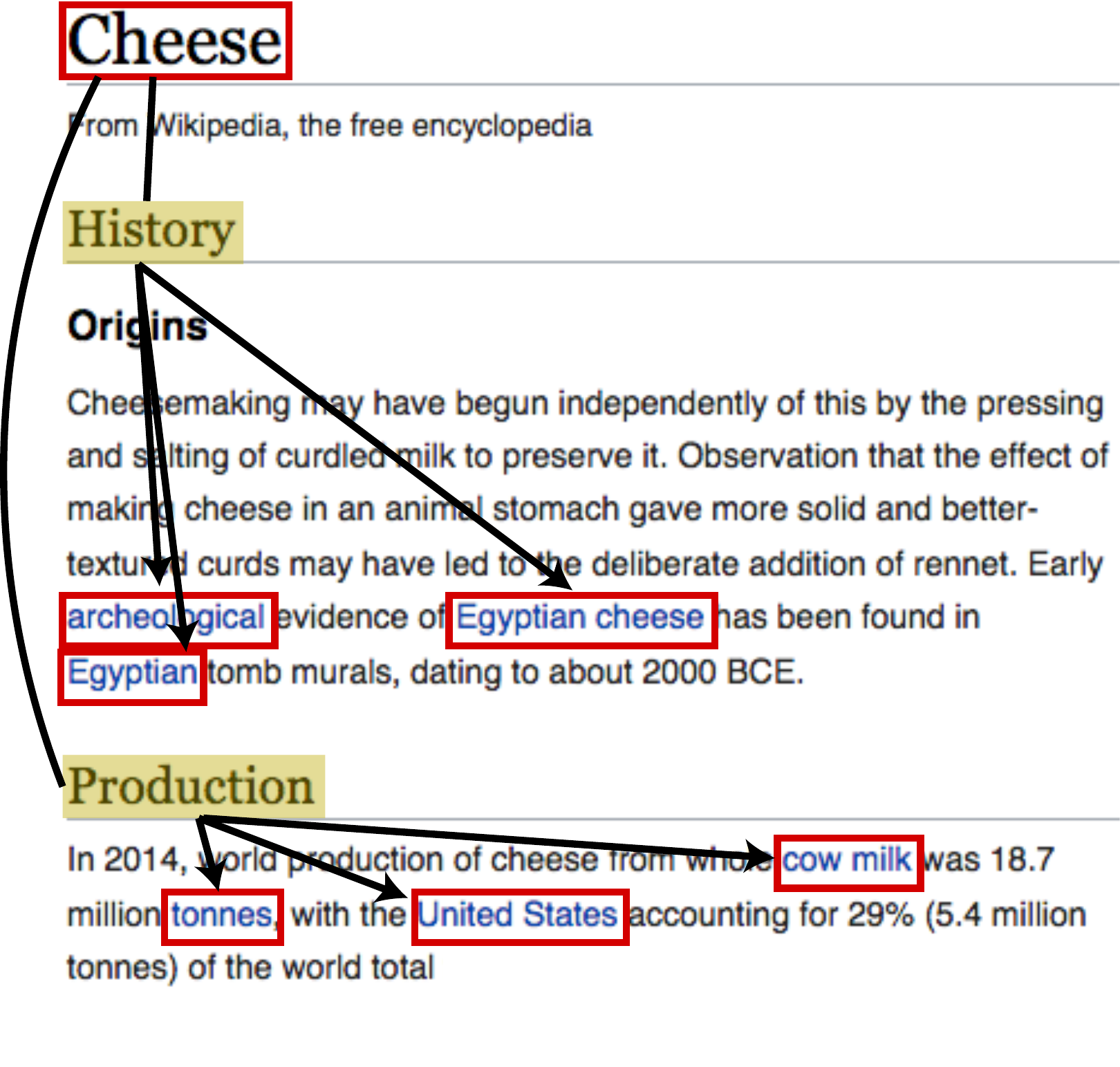}
\caption{Example graph construction strategy from a Wikipedia article excerpt. Red boxes are nodes representing the corresponding entity, and yellow boxes are edge labels. Adapted from \url{https://en.wikipedia.org/wiki/Cheese}.}\label{fig:graph-construction}
\end{figure}

Before we can train knowledge graph embeddings, we must have a knowledge graph. Since the evaluation topics we use come from Wikipedia, we cannot use a freely-available knowledge graph based on all of Wikipedia data (e.g., Freebase or DBPedia)---this could artificially improve results by including information from the relevant paragraphs themselves.\footnote{We also cannot remove the evaluation topics when training the graph because that defeats the purpose; without target entities encoded in the embeddings, there is no way to find similar entities when ranking.} Instead, we construct our own knowledge graph from the available CAR training data. Our approach is based on the observation made by WikiLinks~\cite{singh2012wikilinks}: a knowledge graph can be constructed using links in Wikipeida articles between the article's entity and entities that the page links to. Unlike WikiLinks, we enrich our knowledge graph by labeling the edges using heading information.

More formally, let knowledge graph $G=(E,R)$, where $E$ is the set of entities, and $R$ is the set of relations. Let $E$ be of the union of the set of all article topic entity and the set of all entities found in links. The set of directed relations $R$ is defined as any pair of entities for $(t,m)$ where $t$ is an article topic, and $m$ is an entity mention in a paragraph relevant to the query. The edges are labeled using the highest-level (non-title) heading of the paragraph. An illustration of this process is shown in Figure~\ref{fig:graph-construction}. To address the data sparsity problem caused by the large number of low-frequency headings, we only the lemmatized syntactic head of the main heading. Furthermore, we combine any edge label that doesn't appear in the $e_{max}$ most frequent headings into a single edge label. By using this approach, we are able to encode entity relations in a way that maintains relations between high-frequency headings---precisely the headings that need contextual hints from entities.

We use the knowledge graph $G$ to construct HolE embeddings for the entities and relations. HolE embeddings have been shown to perform the best at entity link prediction by capturing rich interactions via circular correlation~\cite{Nickel2016HolographicEO}. First, we collect all entity mentions in the paragraph being ranked. We collect entities from links that appear in Wikipedia paragraphs that target other articles. However, because Wikipedia guidelines suggest only linking an entity on its first occurrence in an article~\footnote{\url{https://en.wikipedia.org/wiki/Wikipedia:Manual_of_Style/Linking}}, we also explore using an entity extractor to find entity mentions (DBPedia Spotlight~\cite{isem2013daiber}). For each entity mention, we calculate the entity similarity score, given the current query topic entity and the current heading label. We include the top $n_{entscores}$ from the paragraph during the combination phase of the model (see Figure~\ref{fig.model}d). This is similar to how some models perform k-max pooling of query term results. This approach differs from contextual vectors because these signals refer to the entire query-document combination, and not specific query terms.

\subsection{Implementation details}\label{sec:meth-pacrr}
We apply our methods for altering the generalized interaction-focused neural ranking model to the PACRR model~\cite{Hui2017PACRRAP}. This model has been demonstrated to be a strong approach for ad-hoc retrieval. To understand how our methods are integrated into the system, we provide a short description of the model architecture (see~\cite{Hui2017PACRRAP} for full details). For the matching phase, PACRR calculates a $Q\times d$ similarity matrix between query and document terms using word embedding similarity scores. It then performs square convolutional filters of multiple sizes (e.g., $2\times 2$, $3\times 3$, etc.) over the resulting similarity matrix. The result of these filters are max-pooled. Then, scores across query terms are $k$-max pooled ($k=2$) to get the top scores for each query term. Here, the query term results are concatenated with IDF scores for each term. Finally, the results are combined using a simple dense combination layer to produce a final query-document relevance score.

We include contextual vectors after the query term pooling. The vectors are simply concatenated alongside the IDF vector. For heading independence, we split out the work up through the query term pooling. We add an additional dense layer here to further consolidate heading position information before the final combination phase. We include knowledge graph embedding scores in the combination phase, alongside the matching scores.


\section{Experiment}
\label{sec.eval}

In this section, we describe our primary experiment using the approaches described in Section~\ref{sec.method} and present our results using the CAR dataset.

\subsection{Experimental Setup}

\begin{table}
\centering
\caption{Dataset characteristics from the v1.5 data release, including counts of automatic (Auto.) and Manual (Man.) relevance judgments.}
\label{tab_data}
\begin{tabular}{lrrrr}
\toprule
&&&\multicolumn{2}{c}{Relevance} \\ \cmidrule{4-5}
Dataset & Articles & Queries & Auto. & Man. \\
\midrule
\texttt{train.fold1-2} (train) & 114,387 & 873,746 & $2.2M$ & - \\
\texttt{test200} (validation) & 198 & 1,860 & $4.7k$ & - \\
\texttt{benchmarkY1test} (test) & 133 & 2,125 & $5.8k$ & $30k$ \\
\bottomrule
\end{tabular}
\end{table}

\begin{table}
\centering
\caption{Manual relevance judgment counts and occurrence frequencies for benchmarkY1test.}
\label{tab_relevance_counts}
\begin{tabular}{lrrr}
\toprule
Relevance Label & Count & \% & Value \\
\midrule
Must be mentioned & 2,461 & 8.3\% & 3 \\
Should be mentioned & 1,970 & 6.7\% & 2 \\
Can be mentioned & 3,094 & 10.5\% & 1 \\
Roughly on topic & 9,219 & 31.2\% & 0 \\
Non-relevant & 12,785 & 43.2\% & -1 \\
Trash & 42 & 0.1\% & -2 \\
\bottomrule
\end{tabular}
\end{table}

\subsubsection{Dataset}
We use the official TREC CAR dataset (version 1.5) for both training and evaluating our approach~\cite{dietz2017cardata,dietz2017car}. This dataset was constructed from Wikipedia articles that represent topics (that is, it does not include meta or talk pages). All the paragraphs are collected, disassociated with their articles and surrounding content, and used as the main collection for retrieval (30M paragraphs, \texttt{paragraphcorpus}). The dataset also includes queries, which were automatically generated from the article structure. For each query, automatic relevance judgments are provided based on the assumption that paragraphs under a particular heading are all valid answers to the corresponding query.

The dataset is split into subsets suitable for training and testing of systems (summary in Table~\ref{tab_data}). For a randomly-selected half of Wikipedia, all data are provided for training (split into 5 folds, \texttt{train.fold0-4}). We use folds 1 and 2 for training our models. A subset of approximately 200 articles from \texttt{train.fold0} was selected as an evaluation dataset by~\cite{DBLP:journals/corr/NanniMMD17} (\texttt{test200}). We use this as a validation dataset. Finally, 133 articles from the second half of Wikipedia (the half that was not designated for training) were selected as evaluation articles (\texttt{benchmarkY1test}). The TREC CAR 2017 task released manual relevance judgments in addition to the automatic judgments available for the other datasets~\cite{dietz2017car}. The manual relevance judgments are graded on a scale from \textit{Must be mentioned} (3) to \textit{Trash} (-2). A summary of the frequency of these labels is given in Table~\ref{tab_relevance_counts}. While the manual relevance judgments are considered gold standard and are capable of matching relevant paragraphs from other articles, they only cover a subset of queries (702 of the 2,125 queries).

\subsubsection{Knowledge graph embeddings}
We first generate a knowledge graph using the method described in Section~\ref{sec:meth-kg}. We use the entire \texttt{train.fold0-4} dataset to crease as extensive of a graph as possible. We generate two versions of the graph: one using hyperlinks as entity mentions, and one using entity mentions extracted using the automatic entity extractor DBPedia Spotlight~\cite{isem2013daiber} with a confidence setting of 0.5. Although this tool is less accurate than manually-created links, it captures entities that are not linked (e.g., the Wikipedia guidelines suggest only linking the first mention of an entity in an article, leaving out subsequent mentions from the graph). Edge labels are limited to only the $e_{max}=1000$ most frequently-used labels. We test both versions of the graph when evaluating the performance of using knowledge graph embedding scores. We train HolE embeddings using the link graph for 5,000 iterations (we found this to be enough iterations for the training to converge), and choose the embeddings from the epoch with the lowest error rate. When picking which entity scores to include in the model, we use top $n_{nentscores}=2$ similarities (this matches the document term pooling parameter $k$ in PACRR).

\subsubsection{Training and evaluation}
We train and evaluate several variations of the PACRR model to explore the effectiveness of each approach.
We train each model for 80 epochs with samples from \texttt{train.fold1-2} (we found this to be adequate based on pilot experiments). Automatic relevance judgments serve as a source of relevant documents, and we use the top non-relevant BM25 documents as negative training examples. Negative samples are used for the pairwise loss function used to train PACRR, and BM25 results offer higher-quality negative samples than random paragraph would (e.g., these examples have matching terms, whereas random paragraphs likely would not).\footnote{We acknowledge that some paragraphs included as negative training samples, if inspected manually, would be found relevant due to the limitations of the automatic relevance judgments. We deem this as okay, considering the high occurrence of non-relevant documents in the manual relevance judgments, and the comparatively poor performance of BM25 at CAR.} For each positive sample, we include 6 negative samples. To a point, including more negative samples has been shown to improve the performance of PACRR at the expense of training time~\cite{MacAvaney2017AnAF}; we found 6 negative samples to be an effective balance between the two considerations. The training iteration that yields the highest R-Precision value on the validation dataset (\texttt{test200}) is selected for evaluation on the test dataset. We then rerank the top 100 BM25 results for each query in \texttt{benchmarkY1test}, and test using the manual and automatic relevance judgments. For each configuration, we report the 4 official TREC CAR metrics:
Mean Average Precision (MAP), R-Precision (R-Prec), Mean Reciprocal Rank (MRR), and normalized Discounted Cumulative Gain (nDCG). We compare the results to an unmodified version of PACRR trained using the same approach, the initial BM25 ranking, and the other top approaches submitted to TREC.

\begin{table*}
\caption{Performance results on test using manual and automatic relevance judgments. Manual is evaluated both including and excluding documents without relevance judgments. The top value in each section is in bold. Records marked with * were included in the TREC document pool. Significant results compared to the unmodified PACRR model within each section are marked with $\blacktriangle$ and $\blacktriangledown$ (paired t-test, 95\% confidence). The abbreviations for our methods are as follows: HP is the heading position contextual vector; HF is the heading frequency contextual vector; HI is heading independence; KG (links) uses knowledge graph similarity scores using entity mentions from Wikipedia links; KG (extr.) uses knowledge graph similarity scores using entity mentions extracted automatically from the paragraph.}\label{tab:results}
\begin{tabular}{lrrrr}
\toprule
Approach & MAP & R-Prec & MRR & nDCG\\
\midrule
\bf Manual (including unjudged)\\
PACRR (no modification) &  0.208 &  0.219 &  0.445 &  0.403 \\
PACRR + HP* &  0.209 &  0.218 &  0.452 &  0.406 \\
PACRR + HP + HF* & \bf\up 0.211 & \bf 0.221 & \bf 0.453 & \bf\up 0.408 \\
PACRR + HI &  0.205 &  0.213 &  0.442 &  0.403 \\
PACRR + HI + HF & 0.204 &  0.214 &  0.440 &  0.401 \\
PACRR + HI + HF + KG (links) & \dn 0.198 & \dn 0.206 & 0.429 & 0.395 \\
PACRR + HI + HF + KG (extr.) & 0.200 & \dn 0.206 & 0.433 & 0.396 \\
\\
Sequential dependence model*~\cite{cuis2017} & \dn 0.172 & \dn 0.186 & \dn 0.393 & \dn 0.350 \\
Siamese attention network*~\cite{utd2017} & \dn 0.137 & \dn 0.171 & \dn 0.345 & \dn 0.274 \\
BM25 baseline* & \dn 0.138 & \dn 0.158 & \dn 0.317 & \dn 0.296 \\
\midrule
\bf Manual (excluding unjudged)\\
PACRR (no modification)      &      0.471 &      0.496 &      0.774 &      0.583 \\
PACRR + HP*                  &\up   0.479 &\up   0.504 &\up   0.791 &\up   0.590 \\
PACRR + HP + HF*             &\up   0.480 &\up\bf0.509 &\up\bf0.794 &\up   0.592 \\
PACRR + HI                   &\up   0.479 &\up   0.505 &\up   0.795 &\up   0.593 \\
PACRR + HI + HF              &\up   0.479 &      0.499 &      0.782 &\up   0.590 \\
PACRR + HI + HF + KG (links) &\up   0.482 &\up   0.505 &      0.787 &\up   0.592 \\
PACRR + HI + HF + KG (extr.) &\up\bf0.485 &\up   0.505 &\up   0.792 &\up\bf0.594 \\
\\
BM25 baseline*               &\dn   0.452 &\dn   0.476 &\dn   0.747 &\dn   0.566 \\
\midrule
\bf Automatic \\
PACRR (no modification) &  0.164 &  0.131 &  0.247 &  0.254 \\
PACRR + HP* & \up 0.170 &  0.135 & \up 0.258 & \up 0.260 \\
PACRR + HP + HF* & \up 0.170 &  0.134 & \up 0.255 & \up 0.259 \\
PACRR + HI & \up 0.171 &  0.139 & \up 0.256 & \up 0.260 \\
PACRR + HI + HF & \bf\up 0.176 & \bf\up 0.146 & \bf\up 0.263 & \bf\up 0.265 \\
PACRR + HI + HF + KG (links) & \up 0.174 & \up 0.141 & \up 0.260 & \up 0.263 \\
PACRR + HI + HF + KG (extr.) & \up 0.173 & \up 0.140 & \up 0.260 & \up 0.262 \\
\\
Sequential dependence model*~\cite{cuis2017} & \dn 0.150 & \dn 0.116 & \dn 0.226 & \dn 0.238 \\
Siamese attention network*~\cite{utd2017} & \dn 0.121 & \dn 0.096 & \dn 0.185 & \dn 0.175 \\
BM25 baseline* & \dn 0.122 & \dn 0.097 & \dn 0.183 & \dn 0.196 \\
\bottomrule
\end{tabular}
\end{table*}

\subsection{Results}\label{sec.results}

We present the performance of our methods on CAR in Table~\ref{tab:results}. Overall, the results show that our methods perform favorably compared to the unmodified PACRR model, the other other top submissions to TREC CAR 2017 (sequential dependency model~\cite{cuis2017} and the Siamese attention network~\cite{utd2017}), and the BM25 baseline (which our method re-ranks).

Due to the relatively low number of manual relevance judgments per query\footnote{On average, there are 42 manual relevance judgments for the 702 queries that were manually assessed.}, we report manual relevance judgments both including and excluding unjudged paragraphs. When unjudged paragraphs are included, they are assumed to be irrelevant.\footnote{Unjudged evaluation is unavailable for the sequential dependency model and Siamese attention network.} When unjudged documents are excluded, we filter unjudged paragraphs from the ranked lists (i.e., we perform a condensed list evaluation). The condensed list evaluation is a better comparison for methods that were not included in the evaluation pool~\cite{sakai2008information}. Indeed, PACRR with the heading position and heading frequency contextual vectors outperforms the all other approaches, and significantly outperforms the unmodified PACRR model in terms of MAP and nDCG. Interestingly, the knowledge base approaches perform significantly worse than the unmodified PACRR in terms of R-Prec when unjudged paragraphs are included. However, when unjudged paragraphs are not included, it performs significantly better in every case. In fact, the version that uses entity extraction when calculating entity scores performs best overall in terms of MAP and nDCG. This shows that these approaches are effective, and likely rank paragraphs that are relevant yet unjudged high (reducing the score when unjudged documents are included).

When considering automatic relevance judgments, heading independence and the heading frequency contextual vector significantly outperforms unmodified PACRR, and performs best overall in every metric. However, the performance when evaluating with manual judgments does not significantly outperform unmodified PACRR. Since training is also conducted using automatic relevance judgments, this may be caused by PACRR overfitting to this sense of relevance. Interestingly, when the knowledge graphs features are added to this approach, performance drops with automatic judgments, but increases with manual judgments. This suggests that these features are useful for determining human-classified relevance.


\section{Analysis}
\label{sec.analysis}

In this section, we investigate several questions surrounding the results to gain more insights into the behavior and functionality of the approaches to CAR detailed in Section~\ref{sec.method}.

\subsection{Characteristics of heading positions}
Our approaches assert that heading positions (i.e., title, intermediate, and main heading) act as a signal for heading utility. In Section~\ref{sec.results} we showed that including the heading position as either the heading position contextual vector or via and heading independence improves performance beyond the baseline neural architecture. Here, we test the hypothesis that terms found in different heading positions exhibit different behaviors in relevant documents.

We use the \textit{term occurrence rate} to assist in this analysis. We define term occurrence rate $occ(h)$ for a given heading $h$ as the probability that any term in a given heading appears in relevant paragraphs. More formally:
$$
occ(h)=\frac
  {\sum_{p\in rel(h)}{I(\exists t\in p | t \in h)}}
  {|rel(h)|}
$$
where $rel(h)$ is the set of relevant paragraphs for $p$ and $I$ is the indicator function. Although the term occurrence rate only accounts for a single binary term match within relevant documents, this assumption is justified by the fact that headings are usually terse, and the neural architecture we use only considers the top 2 matches for each query term for ranking purposes (PACRR's query term pooling).

\begin{figure}
\scalebox{0.85}{\input{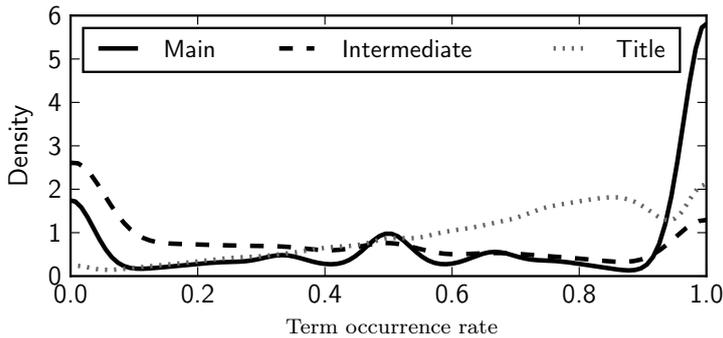}}
{\par\hspace{14em}\vspace{-0.5em} Term occurrence rate}\vspace{1em}
\caption{Kernel density estimation for main (solid), intermediate (dashed), and title (dotted) heading term occurrence rates, based on automatic judgments in train.fold0.}\label{fig.comp}
\end{figure}

In Figure~\ref{fig.comp}, we plot a kernel density estimation of term occurrence distributions for each heading position. We use all topics from the training dataset, and calculate term occurrence rates using automatic relevance judgments. The figure shows that main headings are much more likely to appear in relevant documents than titles and intermediate headings, with a much higher density at the term occurrence rate of 1. This matches our prediction that main headings are more likely to be topical, and therefore appear in relevant documents. Furthermore, the distributions of intermediate and title headings are roughly opposite each other, with titles more likely to occur than intermediate headings. Note that, due to the hierarchical nature of CAR queries, intermediate headings also appear as main headings in other queries for the same topic. This contributes to the high frequency of main headings with a term occurrence rate of 0. Furthermore, many main headings are only used in a single article (i.e., only appear once in the Wikipedia collection), with only a handful of paragraphs associated with them. This explains the multi-modal distribution seen for main headings; the small denominators result in local maxima near 0.33, 0.5, and 0.67.

\subsection{Characteristics of high-frequency headings}
In Section~\ref{sec.results}, we showed that including the heading frequency contextual vector improved retrieval performance. We predicted that the model is able to use this information to predict which query terms are likely to appear in relevant documents. To investigate this behavior, we look at whether heading frequency is correlated with term occurrence rates, and whether there is a performance gap between low- and high-frequency headings.

\begin{figure}
\includegraphics[scale=0.7]{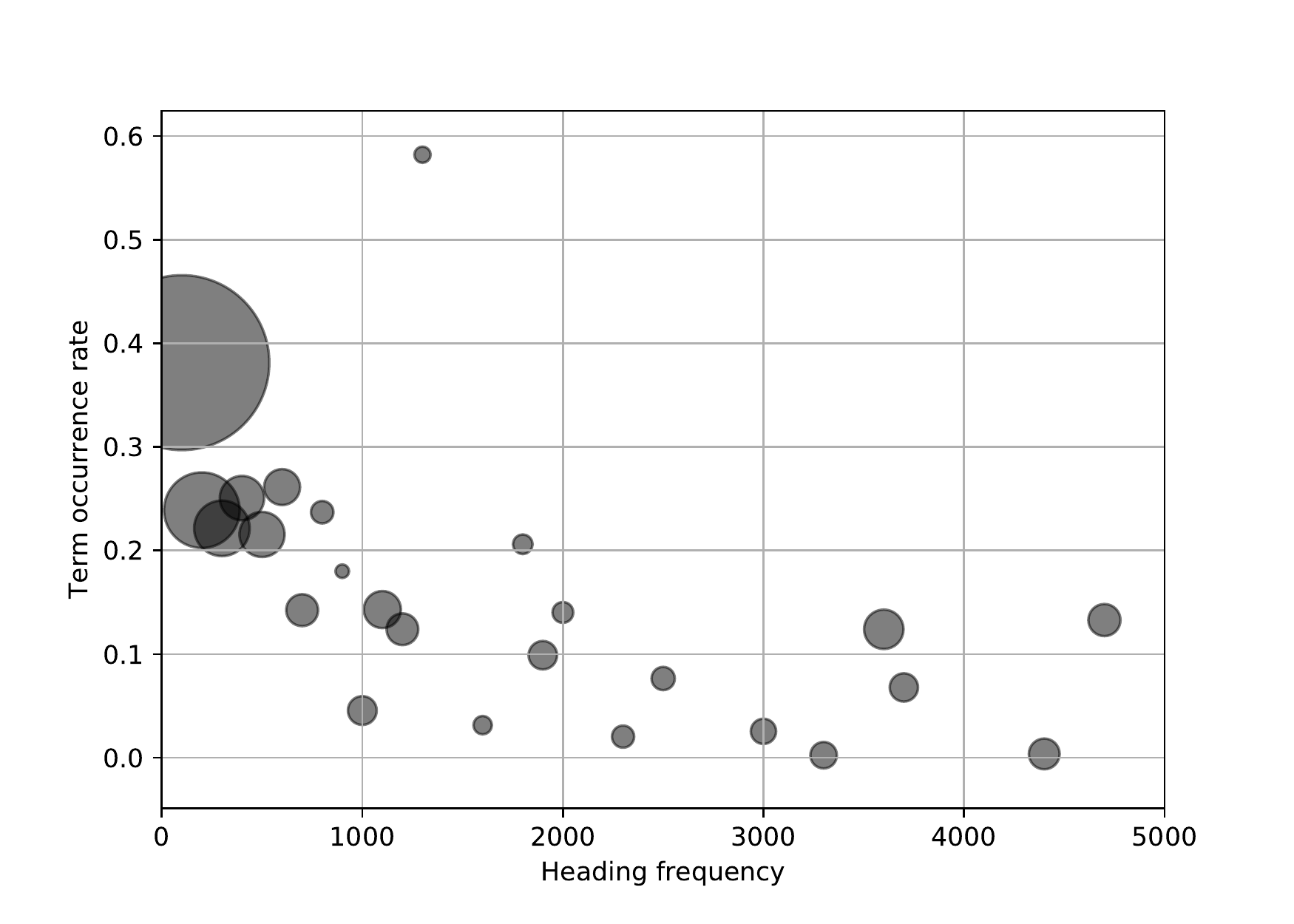}
\caption{Term occurrence rate plotted by heading frequency. Heading frequency is grouped and averaged by 100 for clarity. The area of each point is proportional to the number of heading instances used to calculate the term occurrence rate. One very high frequency heading (History) was omitted for clarity (heading frequency: 15,220, term occurrence rate: 0.035). The trend shows that low-frequency headings are more likely to appear verbatim in relevant paragraphs than high-frequency headings.}\label{fig.frq}
\end{figure}

In Figure~\ref{fig.frq}, we plot term occurrence rates by heading frequency. The figure only includes headings that occur at least twice in \texttt{train.fold0}, and we remove the extremely-frequent heading `History' to improve readability, and points are binned with a size of 100. The area of each point is proportional to the number of heading instances. It is clear from the figure that the less frequent a heading is, the higher the term occurrence rate. In general, headings with a frequency less than 1,000 have an average heading frequency 10-20 points higher than higher-frequency headings.

\begin{table}
\caption{MAP scores stratified by heading frequency of main heading for each query (manual relevance judgments, excluding unjudged). Infrequent headings that do not appear elsewhere in the dataset are included in the `Infrq.' column. The approaches yield the highest performance improvement on the high-frequency (low-utility) queries.}\label{tab.breakdown}
\begin{tabular}{lrrrrrr}
\toprule
Model & Infrq. & 0-20\% & 20-40\% & 40-60\% & 60-80\% & 80-100\% \\
\midrule
PACRR (unmod.)               &   0.503 &   0.521 &   0.453 &   0.388 &   0.412 &   0.366 \\
\hspace{1em}+ HP             &   0.509 &   0.520 &   0.464 &   0.400 &\bf0.430 &   0.377 \\
\hspace{1em}+ HP + HF        &   0.511 &   0.519 &   0.470 &   0.406 &   0.417 &   0.383 \\
\hspace{1em}+ HI             &   0.509 &   0.525 &\bf0.479 &   0.382 &   0.417 &\bf0.391 \\
\hspace{1em}+ HI + HF        &   0.509 &   0.519 &   0.474 &   0.403 &   0.409 &   0.379 \\
\hspace{1em}+ HI + HF + KG (links) &   0.511 &   0.529 &   0.475 &   0.412 &   0.408 &   0.376 \\
\hspace{1em}+ HI + HF + KG (extr.) &\bf0.513 &\bf0.535 &   0.477 &\bf0.416 &   0.407 &   0.380 \\
\\
BM25                         &   0.482 &   0.480 &   0.429 &   0.378 &   0.420 &   0.373 \\
\midrule
Query count                  & 349 & 89 & 96 & 81 & 43 & 44 \\
Max. \% increase (to unmod.)& +2.0\% & +2.7\% & +5.7\% & +7.2\% & +4.3\% & +6.8\% \\
\bottomrule
\end{tabular}
\end{table}

\subsection{Difficultly matching high-frequency headings}

Given the knowledge that high-frequency headings generally exhibit a low term occurrence rate, we are interested in measuring whether there exists a discrepancy between the performance at different heading frequencies. Table~\ref{tab.breakdown} shows a MAP~\footnote{We observed similar behavior for MAP, R-Prec, MRR, and nDCG, so we only report MAP here.} performance breakdown on the test dataset (manual relevance judgments, unjudged excluded) stratified by the heading frequency of the main heading found in the training dataset. The number of queries in each stratum varies because the strata were selected from equally-spaced breakpoints in the training data, and the test set does not represent a uniform selection of Wikipedia articles. The most frequent headings (80-100\%, e.g., \textit{`History'} and \textit{`Early Life'}) and exhibit the worst performance among all strata. This matches our intuition that these low-utility headings are difficult to match. The low frequency headings exhibit the highest performance by all models. Notice that the gains compared to the unmodified PACRR are higher for high-frequency headings than the low frequency headings, despite having a worse performance overall. This validates our claims that low-utility headings are harder to match than high-utility headings, and that our approaches are able to improve performance for these queries.


\section{Conclusion} 
\label{sec.conclusion}

In this work, we proposed that a central challenge to CAR is the identification and mitigation of low-utility question facets. We introduced two techniques for identifying low-utility facets: contextual vectors based on the hierarchical structure of CAR questions and corpus-wide facet usage information; and incorporation of independent matching pipelines for separate query components. We then introduced one approach to mitigate low-utility facets, which involves building a knowledge graph from the CAR training data, and using entity similarity scores during query processing. We applied these approaches to a leading neural ranking method, and evaluated using the TREC CAR dataset. We found that our approach improved performance when compared to the unmodified version of the ranker, and improve performance by up to 26\% compared to the next best approach. We then performed an analysis that verified that our indicators of facet utility were valuable, and that our approaches improve the performance on low-utility headings. As one of the first comprehensive works on CAR, we expect our findings to shape the directions taken for CAR in the future.



\bibliographystyle{spbasic}      
\bibliography{bibliography}   

%
%

\end{document}